\documentstyle[12pt,epsfig]{article}

\begin{document}

\begin{center}%
  {\large \bf Directed, Elliptic and Triangular Flows in Asymmetric
    Heavy Ion Collisions }

  \vspace{1.0cm}  
  
  {\bf M.~Bleicher$^{1,3}$, K.~A.~Bugaev$^2$, P.~Rau$^{1,3}$,
    A.~S.~Sorin$^4$, J.~Steinheimer$^{1,3}$, and H.~St\"ocker$^{1,5}$ }

  \vspace{1.cm}
  
  $^1$Frankfurt Institute for Advanced Studies (FIAS), 60438
  Frankfurt, Germany\\
  $^2$Bogolyubov Institute for Theoretical Physics,
  National Academy of Sciences of Ukraine,  03680 Kiev, Ukraine \\
  $^3$Institut f\"ur Theoretische Physik, Goethe-Universit\"at,
  60438 Frankfurt, Germany\\
  $^4$Joint Institute for Nuclear Research (JINR), 141980 Dubna, Moscow
  Region, Russia\\
  $^5$GSI Helmholtzzentrum fur Schwerionenforschung, 64291 Darmstadt,
  Germany\\

  \vspace{1.0cm}
  
  {\small emails: \quad bleicher@th.physik.uni-frankfurt.de \quad
    bugaev@th.physik.uni-frankfurt.de \quad
    rau@th.physik.uni-frankfurt.de \quad sorin@theor.jinr.ru \quad
    steinheimer@th.physik.uni-frankfurt.de \quad h.stoecker@gsi.de}

  \vspace{1.cm}  
\end{center}

\abstract{In this paper we propose to thoroughly investigate
  asymmetric nuclear collisions both in the fixed target mode at the
  laboratory energy below 5 GeV per nucleon and in the collider mode
  with a center of mass energy below 11 GeV per nucleon.  Using the
  UrQMD transport model, we demonstrate a strong enhancement of
  directed and elliptic flow coefficients for the midcentral
  asymmetric nuclear collisions compared to symmetric collisions.  We
  argue that such an enhancement is due to the disappearance of the
  nuclear shadowing effect on the side of the smaller projectile
  nucleus.  An analysis of the energy and centrality dependencies of
  the directed, elliptic and triangular flows at midrapidity shows us
  their sensitivity to the details of the employed model of hadronic
  interaction.  In general, the flow patters found for asymmetric
  nuclear collisions have a very rich and complicated structure of
  energy and centrality dependencies compared to the flows found for
  symmetric collisions and are worth to be investigated
  experimentally.  The directed, elliptic, and triangular flow
  coefficients are computed for target nuclei containing high density
  fluctuations and thoroughly compared with that ones obtained in the
  absence of such fluctuations.\\

   }



\vspace*{2.0cm}

{\bf 1. Introduction.}~The experimental study of strongly interacting
matter has reached a decisive moment: it is hoped that the low energy
heavy ion collisions programs performed at the CERN SPS and the BNL
RHIC and two new programs that are planned to begin in a few years at
NICA (JINR, Dubna) and FAIR (GSI, Darmstadt) will allow the heavy ion
community to locate the mixed phase of the deconfinement phase
transition and to discover a possible (tri)critical endpoint.  The
major experimental information \cite{QM11:Phenix, QM11:Star,
  QM11:ALICE,QM11:ATLAS, QM11:CMS} is provided by the measurements of
particle yields, one particle momentum spectra, two particle
correlations, and the Fourier components of the collective hadronic
flow \cite{Voloshin:96} known as $v_1$-coefficient (directed flow),
$v_2$-coefficient (elliptic flow), and $v_3$-coefficient (triangular
flow). Although the great success of experiments at the BNL RHIC
\cite{QM11:Phenix, QM11:Star} and at the CERN LHC
\cite{QM11:ALICE,QM11:ATLAS, QM11:CMS} proved the high efficiency of
modern experimental methods, their results also clearly demonstrated
that the heavy ion collisions programs at RHIC, SPS, NICA and FAIR
energy range are not simpler and they require further development of
both new experimental approaches and far more sophisticated
theoretical models in order to reach their goals.  Therefore, in view
of new opportunities opening with the Nuclotron program Baryonic
Matter@Nuclotron (BMN) \cite{NICA, WhitePaper} which will start at
JINR (Dubna) in 2014, we would like to address here some new physical
issues that can be studied at laboratory energies of $2-5$~AGeV for a
wide range of colliding nuclei in the fixed target mode. In future,
they can also be further investigated at the accelerators of new
generation like NICA (JINR, Dubna) and FAIR (GSI, Darmstadt) at a
center of mass energy up to 11~AGeV.

This range of energies was thoroughly investigated in the past at the
GSI SIS and the BNL AGS experiments, but only for symmetric nuclear
collisions A+A.  In this paper we demonstrate by using the
Ultrarelativistic Quantum Molecular Dynamics transport model
UrQMD~\cite{Bass:1998ca,Bleicher:1999xi} that one of the major tasks
of the low energy programs at JINR and GSI could be a systematic study
of directed, elliptic, and triangular flows for non-central asymmetric
nuclear collisions (ANC), i.e.\ for non-central $A + B$ reactions with
$1 \ll B \ll A$ (see Fig.~1). This aspect of heavy ion physics was not
yet systematically explored. Until today there were only a few works
reported for ANC \cite{Baumgardt:1975qv,ANC:1,ANC:2}. Moreover, these
works were completed before a systematic investigation of the Fourier
components \cite{Voloshin:96} of hadronic flow was proposed.
Therefore, here we argue that at lab.\ energies of about $2-10$~AGeV a
systematic study of $v_1$, $v_2$ and $v_3$ Fourier-coefficients of the
azimuthal particle distributions measured in non-central ANC with a
special choice of impact parameter may help to essentially improve our
understanding of the hadronic matter equation of state at high
densities. In addition, under certain conditions, highest baryonic
charge densities can be reached in these
experiments~\cite{ConicWave:10}, what could help clarifying the
question, whether the mixed quark-hadron phase \cite{NICA,WhitePaper,
  Randrup:WP, Quarkyonic0} or the predicted chiral quarkyonic phase
\cite{Quarkyonic1, Quarkyonic2, Quarkyonic3, Quarkyonic0} can be
formed in this energy range.

First theoretical predictions in this energy region were made very
recently \cite{ConicWave:10} and they indicate a very rich picture of
the physical phenomena in ANC to be investigated, e.g.\ the formation
of high density Mach shock waves and their impact on the spatial
distribution of the produced particles and heavier clusters. The main
purpose of this work is to formulate the most promising physical
issues for the ANC of the BMN program at JINR and to work out the
strategy of their experimental exploration.

This work is organized as follows.  In the next section we discuss the
difference of the directed, elliptic, and triangular flows obtained in
ANC and in symmetric nuclear collisions. The third section is devoted
to the analysis of the directed, elliptic and triangular flows that
may develop, if high density fluctuations occur inside the target
nucleus. The last section contains our conclusions.

\begin{figure}[ht]
  \centerline{\psfig{figure=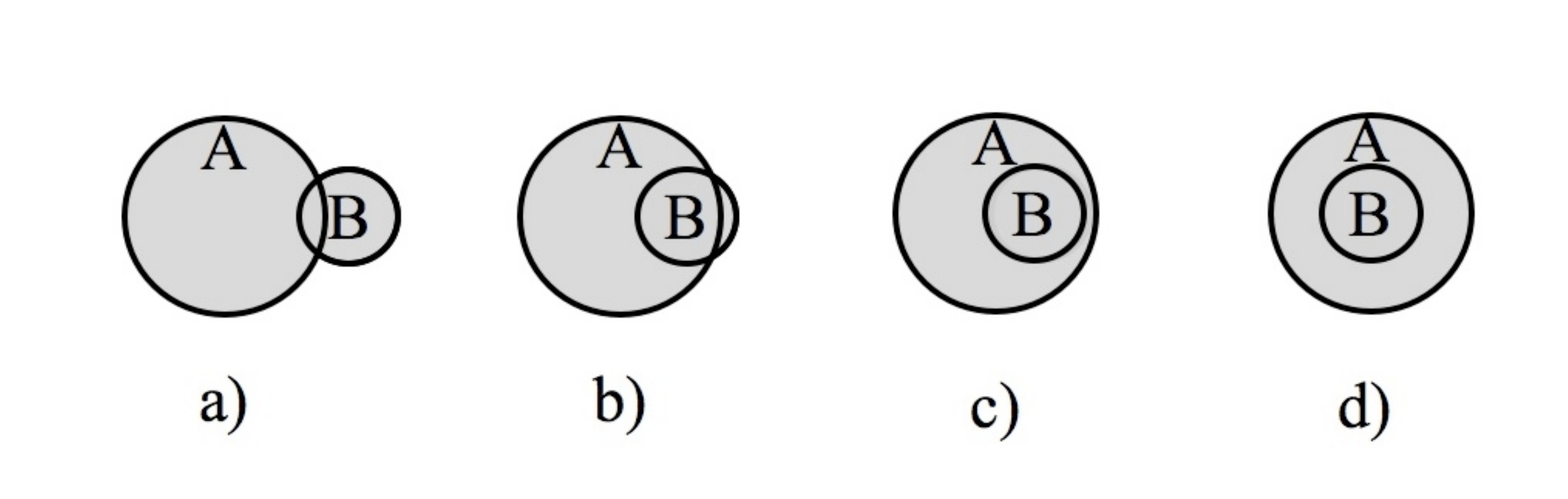,height=5.8cm,width=18.cm}}
  
  {\bf Fig.~1.}  Schematic picture of an asymmetric nuclear collision
  of nuclei A (large circle) with B (small circle) shown in the
  transverse plane.  The common area of two nuclei is shown for very
  peripheral collisions (panel a)), for semi-peripheral collisions
  (panels b) and c)) and for the most central collision (panel d)).
  Clearly, for semi-peripheral collisions the effect of the shadowing
  of particle motion to the right hand side of the nucleus B is very
  weak.
  \label{fig1}
\end{figure}


{\bf 2. The unusual properties of ANC.}~ANC have some history since
they were suggested long ago (see, for instance,
\cite{Baumgardt:1975qv,Hofmann:1976dy,ANC:T79,ANC:T80,Stocker:1981zz}),
but in those days the analysis of Fourier coefficients of the
azimuthal particle distributions was not even suggested. Since the
experiments on collisions of identical nuclei is simpler, ANC were
forgotten for awhile.  After the first work \cite{Voloshin:96} on the
analysis of the Fourier coefficients $v_1, v_2, \ldots$ was published
this method has become a powerful tool for experimental studies of the
evolution process of symmetric heavy ion collisions at high energies
\cite{Residorf:97,Ollitrault, Herrmann,Stock:05, V2scaling}, i.e.\ of
two identical nuclei ($A+A$).  Thus, the quark scaling of the $v_2$
dependence on the transverse momentum $p_T$ \cite{V2scaling} clearly
demonstrated the partonic source of elliptic flow of hadrons at RHIC
energies, while the triangular flow is reflecting the correlations
that appear at the early stage of collisions \cite{Alver:10}.

A principally new element of the ANC compared to symmetric nuclear
collisions is the generation of strong and asymmetric gradients of
energy density and baryonic density at the initial stage of the
collision process along with the stronger flow from the target in the
direction of the projectile nucleus for a specific choice of the
impact parameter values.  In contrast to symmetric collisions, in ANC
the reflection symmetry between the left nucleus A and the right
nucleus B (see panels a)-c) in Fig.~1) is broken from the very
beginning. This leads to an entirely different shape and location of
the overlap region between the colliding nuclei which, in its turn,
results in different flow patterns compared to symmetric $A+A$
collisions.  Indeed, if the impact parameter value is close to
$b_{ANC} \approx 1.1\,{\rm fm}\, \cdot (A^\frac{1}{3} - B^\frac{1}{3})
\pm 1$~fm (see the panels b) and c) of Fig.~1) and if the size of the
target nucleus is chosen close to $B^\frac{1}{3} \approx \frac{1}{2}
A^\frac{1}{3} $, then there is sufficient room to vary the impact
parameter in the experiment and to select values which are close to
$b_{ANC}$ using an event-by-event analysis.  In this case ANC allow us
to scan the interior of the target nucleus, as well as to study in
detail the variety of surface phenomena such as the surface formation
of light nuclear fragments like deuterons, tritons, and heliums
nuclei, the emission of high $p_T$ pions \cite{AntiFlow:Bravina}, the
isospin dependence of hadron surface emission, conical emission due to
Mach shock wave formation in the target~\cite{ConicWave:10}, and so
on.

Consider the change in the directed flow first.  Although the
approximate dependence of the so-called flow parameter $F$
\cite{FlowP} on $A^\frac{1}{3} + B^\frac{1}{3}$, i.e.\ on the
collision time, was found in the symmetric nuclear collision
experiments \cite{SNC:3}, however, a similar dependence of the
directed flow in the ANC is much less certain \cite{Herrmann} and has
to be studied.  In addition, in symmetric nuclear collisions with a
lab.\ energy between 1 and 10~AGeV the effect of nuclear shadowing
plays an important role and causes the negative value of pionic $v_1$
\cite{Ollitrault}.  Since both the peripheral and the most central ANC
(see panels a) and d) of Fig.~1) are very similar to the corresponding
symmetric nuclear collisions due to a similar geometry, one can expect
a similar behavior of their $v_1$ coefficients, i.e.\ $v_1^{AS}
\approx v_1^S $, while for the semi-peripheral ANC (see panels b) and
c) in Fig.~1) one expects a different situation.  This expectation is
supported by the experimental analysis of \cite{ANC:1} of pionic flow
directed to the in-plane $OX$-axis and the one directed oppositely.
Therefore, the semi-peripheral ANC would allow one to essentially
reduce the effect of shadowing on one side of the formed fireball what
could help to finally clarify the source of the strong antiflow of
pions \cite{AntiFlow:Bravina,Stoecker:2004qu} with $p_T < 500$~MeV
observed at SIS energies \cite{AntiFlow:SIS} and predicted earlier
\cite{AntiFlow:Bertsch,AntiFlow:Horst}.  Moreover, one can hope to
study a rich structure of shock waves in hadronic matter (from conical
emission to viscosity) occurring during semi-peripheral ANC
\cite{ConicWave:10}.

In order to demonstrate the new possibilities of the ANC we performed
the analysis of $v_1^{AS}$ using the UrQMD model. This model is one of
the most successful transport models and was supplemented by in-medium
potentials \cite{Li:2005gfa,Li:2006ez,Bleicher} in order to reproduce
the full complexity of the flow patterns in the low energy range.  For
this work, we use the UrQMD model developed in
\cite{Bass:1998ca,Bleicher:1999xi,Li:2005gfa,Li:2006ez,Bleicher} which
was thoroughly tested on the available data in a wide range of
collision energies.  Fig.~2 shows the energy and centrality dependence
of the $v_1^{AS}$ coefficient for all charged particles from the ANC
Ne+Au at longitudinal midrapidity in the equal velocity frame of the
colliding nuclei. Such a system is convenient to compare the results
with both the data obtained in symmetric nuclear collisions and with
UrQMD simulations reported earlier \cite{Bleicher}.  As one can see
from Fig.~2 the $v_1^{AS}$ coefficient of charged particles is
essentially non-zero at $|y| < 0.1$ for ANC, whereas it is zero for $y
= 0$ in symmetric nuclear collision since it is an odd function of the
center of mass rapidity. Also Fig.~2 demonstrates that that inclusion
of the potentials in the UrQMD
model~\cite{Li:2005gfa,Li:2006ez,Bleicher} is important since their
absence leads to roughly double decrease of the directed flow
coefficient.  A comparison of the charged particles directed flow with
that one of positive pions shows a complete similarity in the energy
and centrality dependence (see Fig.~3). The behavior of negative pions
is also similar.  Figs.~2 and 3 demonstrate a gradual decrease of the
directed flow (left-right asymmetry) with the increase of the
collision energy.


\begin{figure}[t]
  
  \centerline{\psfig{figure=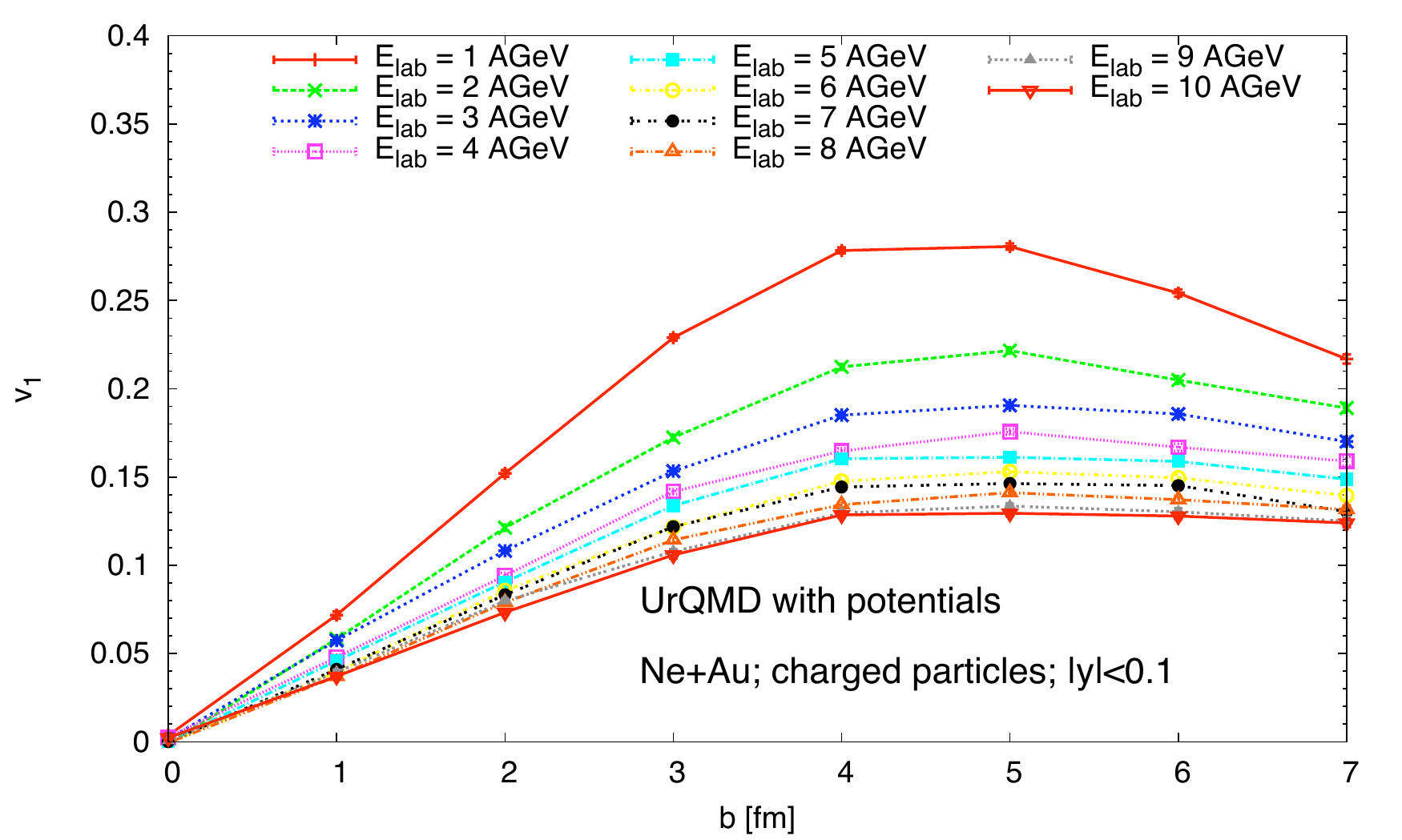,height=5.7cm,width=8.4cm}
    \hspace*{-0.5cm}
    \psfig{figure=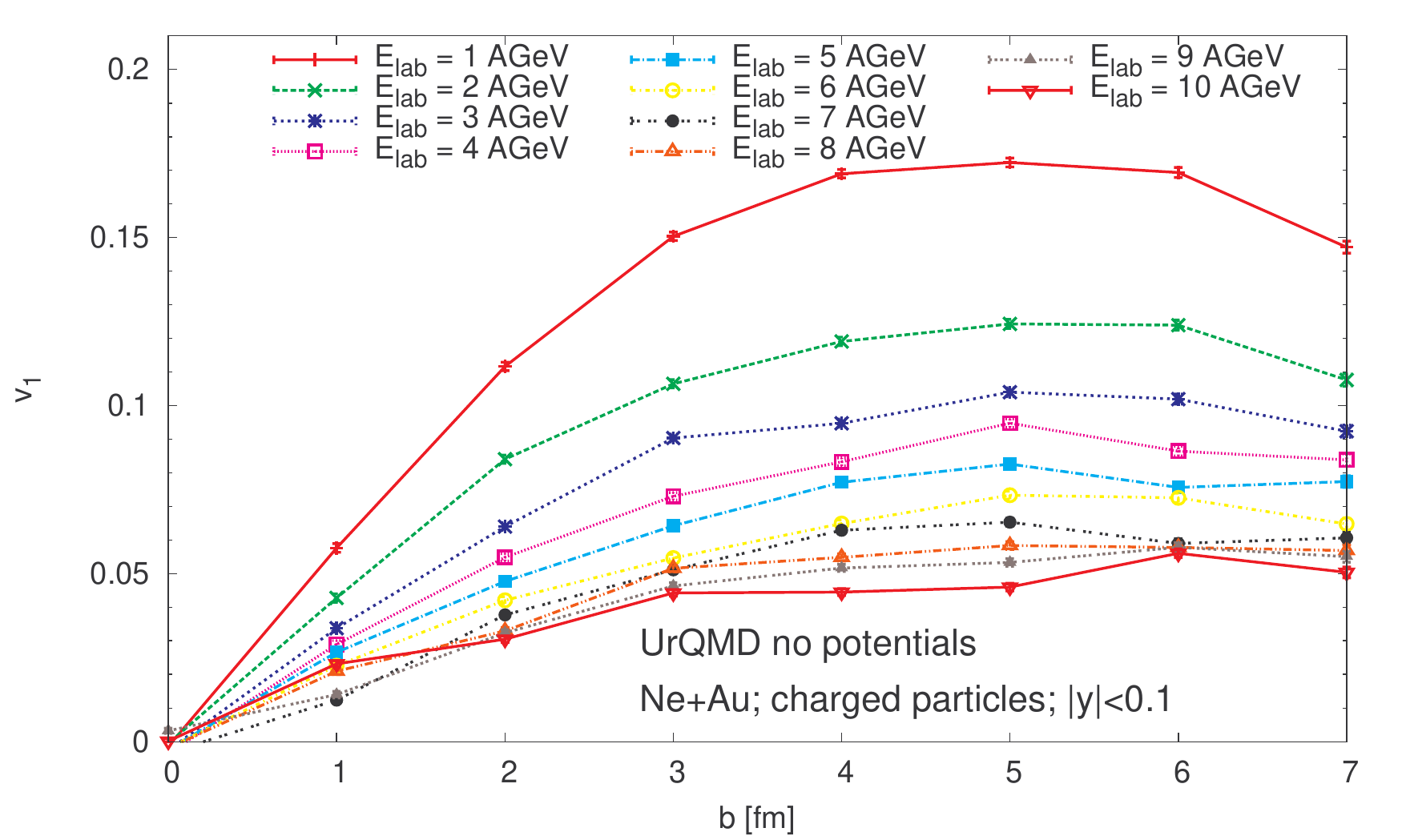,height=5.7cm,width=8.4cm}
  }

  \vspace*{-.05cm}
  
  {\bf Fig.~2.}  Energy and centrality dependence of the $v_1^{AS}$
  coefficient of charged particles from Ne+Au collisions as calculated
  with the UrQMD model with potentials (left panel) and without them
  (right panel). For more details see the text.

\end{figure}\label{fig2}


\begin{figure}[t]
  
  \centerline{\psfig{figure=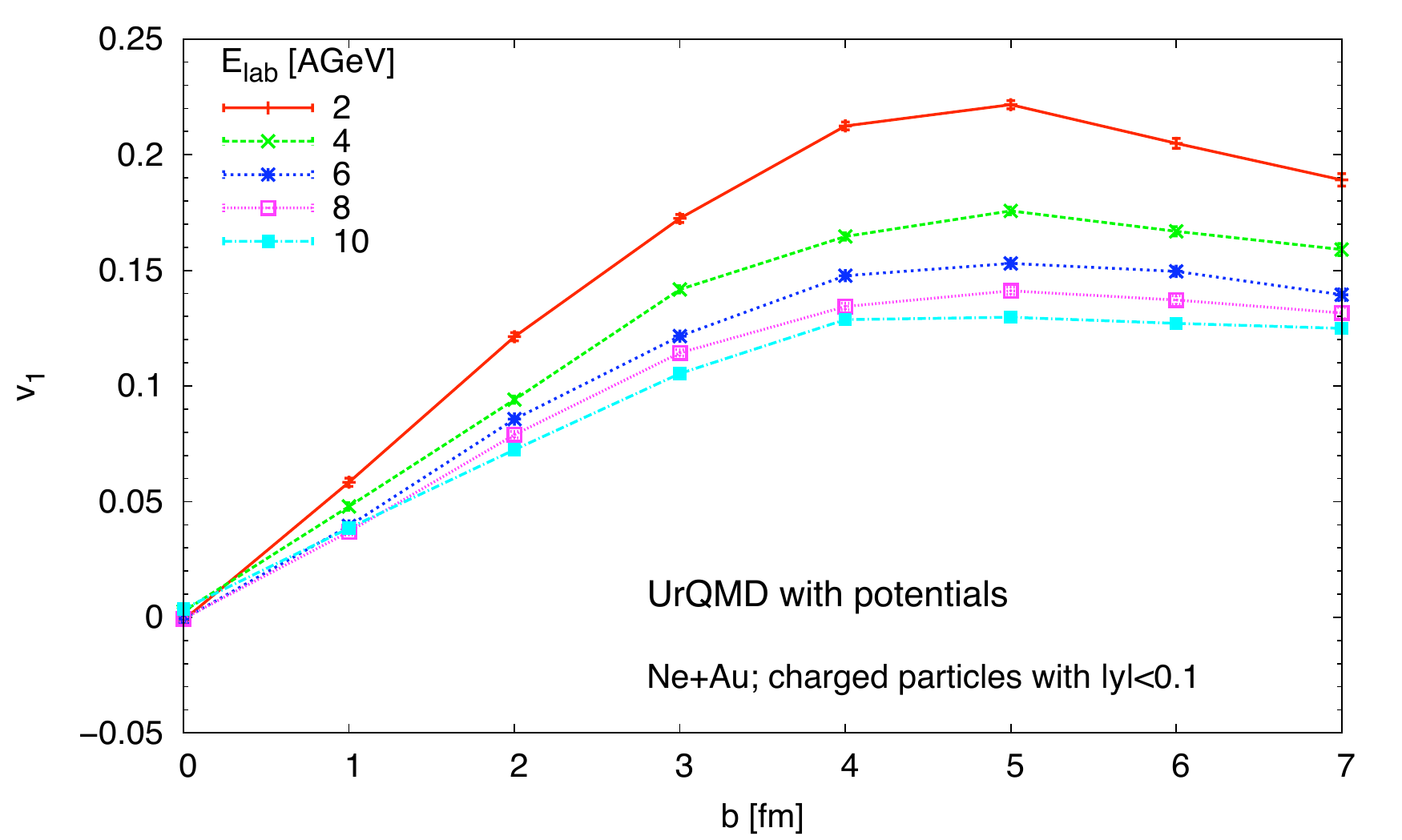,height=5.7cm,width=8.4cm}
    \hspace*{-0.5cm}
    \psfig{figure=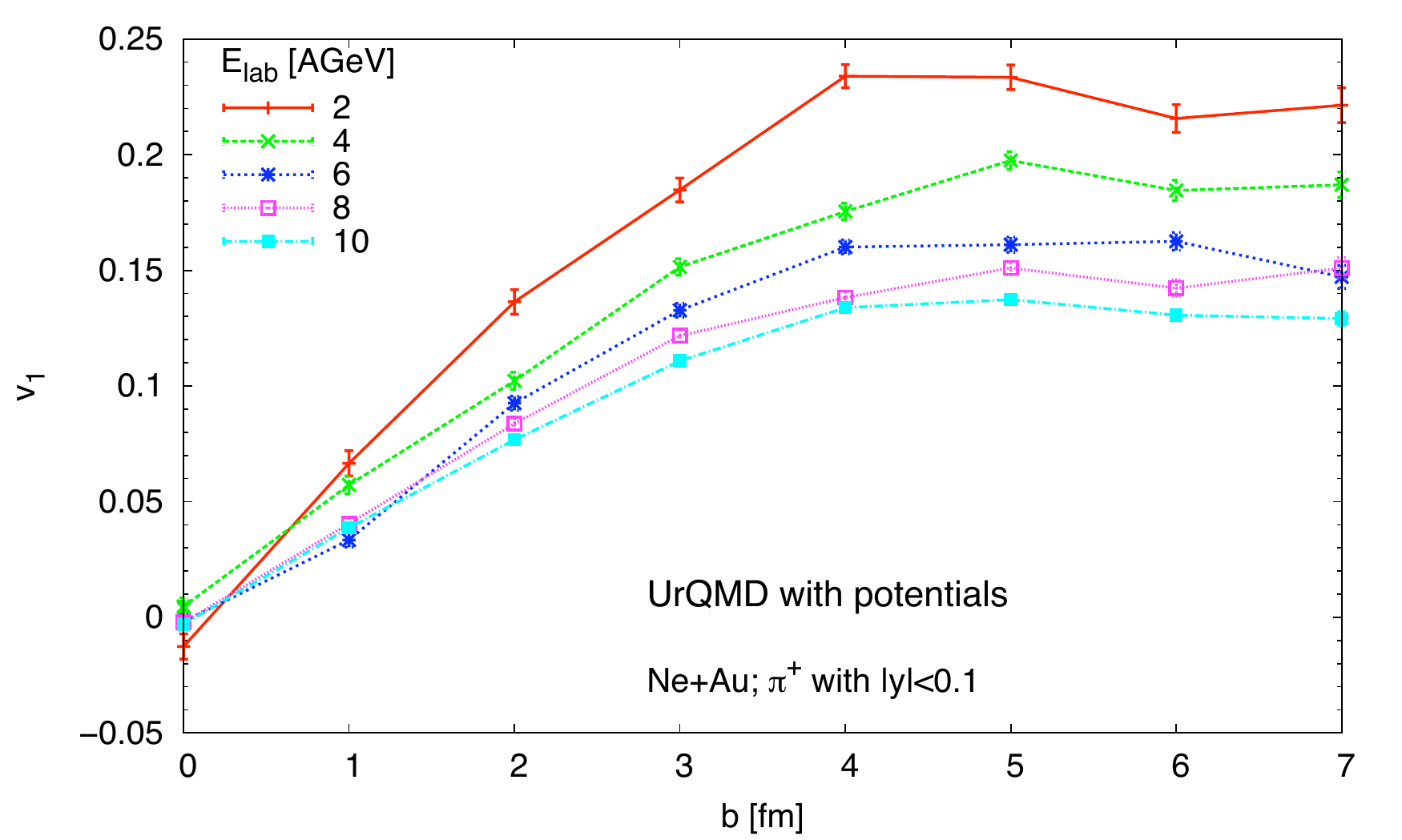,height=5.7cm,width=8.4cm} }

  \vspace*{-.05cm}
  
  {\bf Fig.~3.}  Energy and centrality dependence of the $v_1^{AS}$
  coefficient found by the UrQMD model with potentials for all charged
  particles (left panel) and for positive pions (right panel) for the
  ANC Ne+Au.
  
\end{figure}\label{fig3}


\begin{figure}[ht]
  
  \centerline{\psfig{figure=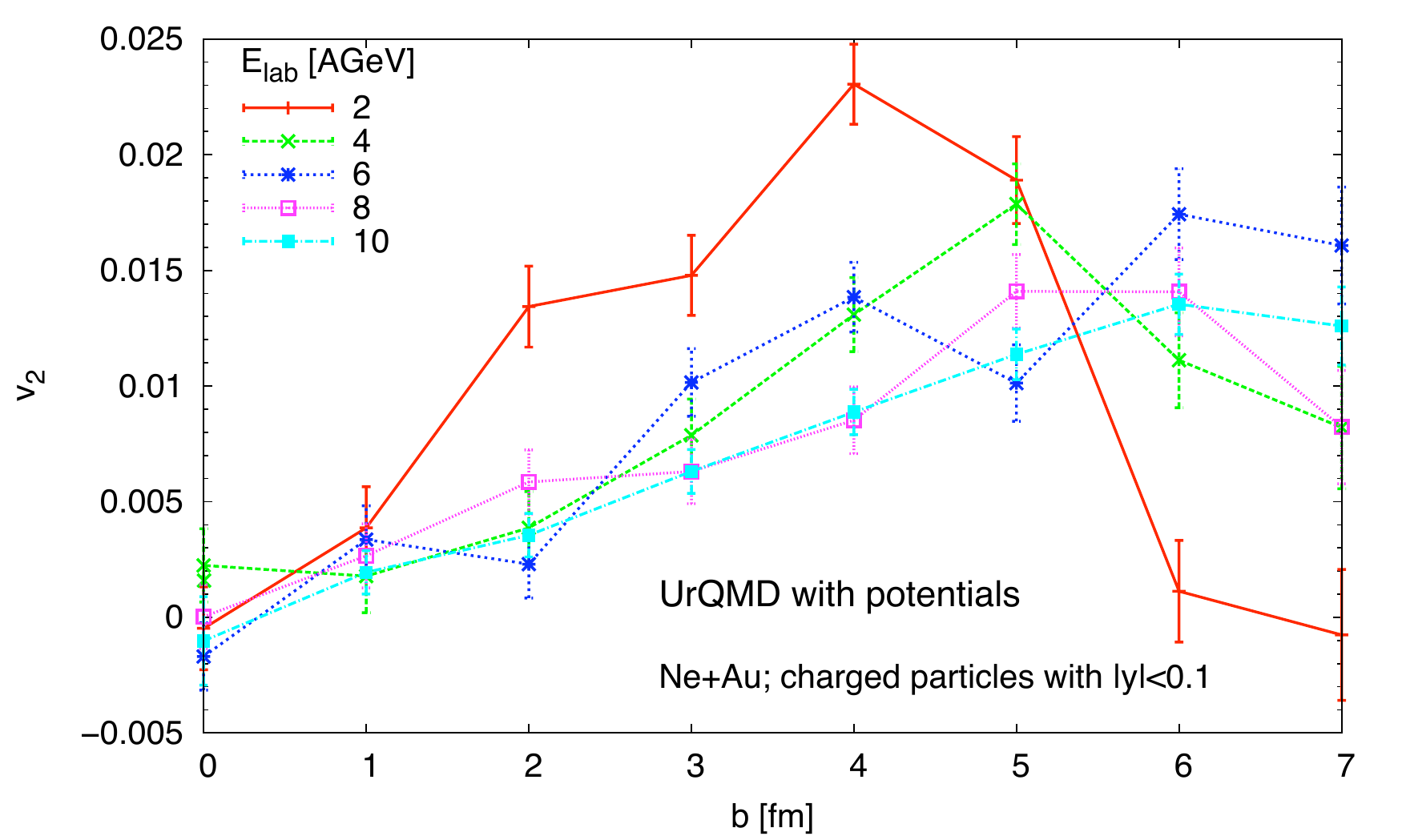,height=5.7cm,width=8.4cm}
    \hspace*{-0.5cm}
    \psfig{figure=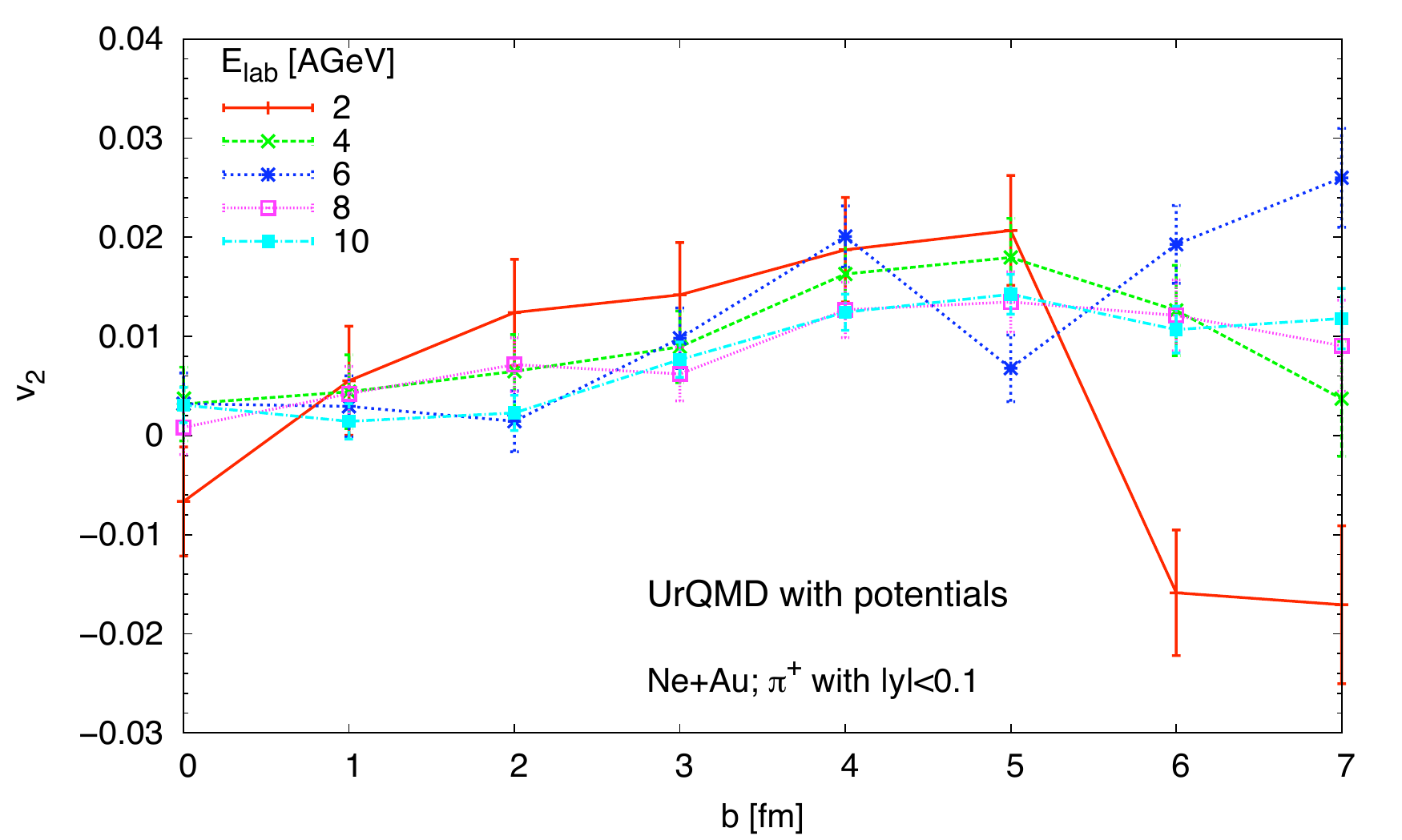,height=5.7cm,width=8.4cm} }

  \vspace*{-.05cm}
  
  {\bf Fig.~4.}  Energy and centrality dependence of the $v_2^{AS}$
  coefficient found by the UrQMD model with potentials for charged
  particles (left panel) and for positive pions (right panel) in the
  ANC Ne+Au.
  
\end{figure}\label{fig4}

Let us now consider the qualitative change of the elliptic flow
(in-plane) coefficient behavior in ANC.  The experimental data show
that at lab.\ energies of about $3.3$~AGeV the elliptic flow in
symmetric collisions is zero, i.e.\ $v_2^S = 0$, and its value slowly
grows with the colliding energy \cite{Stock:05}. The reason for those
small values of the elliptic flow around this energy is that the slow
moving remnants of the target (projectile) nucleus $A$ ($B$) shadow
the motion of particles from the collision zone directed from nucleus
$A$ to nucleus $B$ (see Fig.~1).  The change of the $v_2^S$ sign at
$3.3$~AGeV signifies the transition from the out-of-plane to the
in-plane elliptic flow which was predicted long ago
\cite{Ollitrault:v2}.

Due to a similar geometry both the peripheral and the most central ANC
are alike the corresponding symmetric nuclear collisions, and
therefore, in this case one has to expect almost identical behavior of
their $v_2$ coefficients, i.e.\ $v_2^{AS} \approx v_2^S$.  On the
other hand, the semi-peripheral collisions (panels b) and c) in
Fig.~1) produce an entirely different situation: in this case, if the
smaller nucleus $B$ hits the larger nucleus $A$ at its boundary, not
too far above or below its surface, i.e for impact parameter values
close to $b_{ANC}$, then the whole situation concerning the shadowing
is changed. At these impact parameter values there is almost no
shadowing on one side and, hence, one can expect that in those ANC the
$v_2^{AS}$ coefficient can essentially be enhanced at lab.\ energies
between 2 and 5~AGeV. In addition, one can expect that some number of
slow moving particles with an initial momentum directed from nucleus B
to nucleus A is reflected backwards and creates an additional flow in
the direction of the no-shadowing side of the collision region leading
to positive values of $v_2^{AS}$. These new features of the ANC
elliptic flow can be seen in Fig.~4.  For $E_{\rm lab} = 2 $ AGeV both the
charged particles and pions have positive $v_2^{AS}$ coefficient for
semi-peripheral collisions, while for central and peripheral
collisions the pionic $v_2^{AS}$ is negative.  As the collision energy
is increased, the maximum of the $v_2^{AS}$ coefficient gradually
moves to larger impact parameter values and gets less pronounced.

In this work we report the first simulations of the triangular flow
coefficient for the low collision energy range.  Recent analyses
\cite{Alver:10,QM11:Phenix,QM11:Star,QM11:ALICE,QM11:ATLAS,QM11:CMS,Bass:10}
clearly demonstrated an importance of the $v_3$ coefficient for an
elucidation of the experimental information and here we confirm this
fact.  Although the $v_3^{AS}$ signal found for $E_{\rm lab} = 2 - 10$
AGeV is much weaker than that one determined at RHIC energies, the
structure of triangular flow at low collision energies is definitely
much richer as one can see from Fig.~5.  Indeed, for $E_{\rm lab} = 2 -
3$~AGeV the charge particles (negative pions) predict that the
$v_3^{AS}$ coefficient has a maximum in semi-peripheral collisions
whereas $v_3^{AS}$ is negative at impact parameter values of $b
\approx 1$~fm ($b < 1$ fm) and $b \ge 6 $ fm.  For $E_{\rm lab} =
4-5$~AGeV the maximum of the $v_3^{AS}$ coefficient is still clearly
seen, although it gets wider, whereas for larger collision energies
this maximum is not that clearly visible especially for pions (see the
right panel in Fig.~5).


\begin{figure}[ht]
  
  \centerline{\psfig{figure=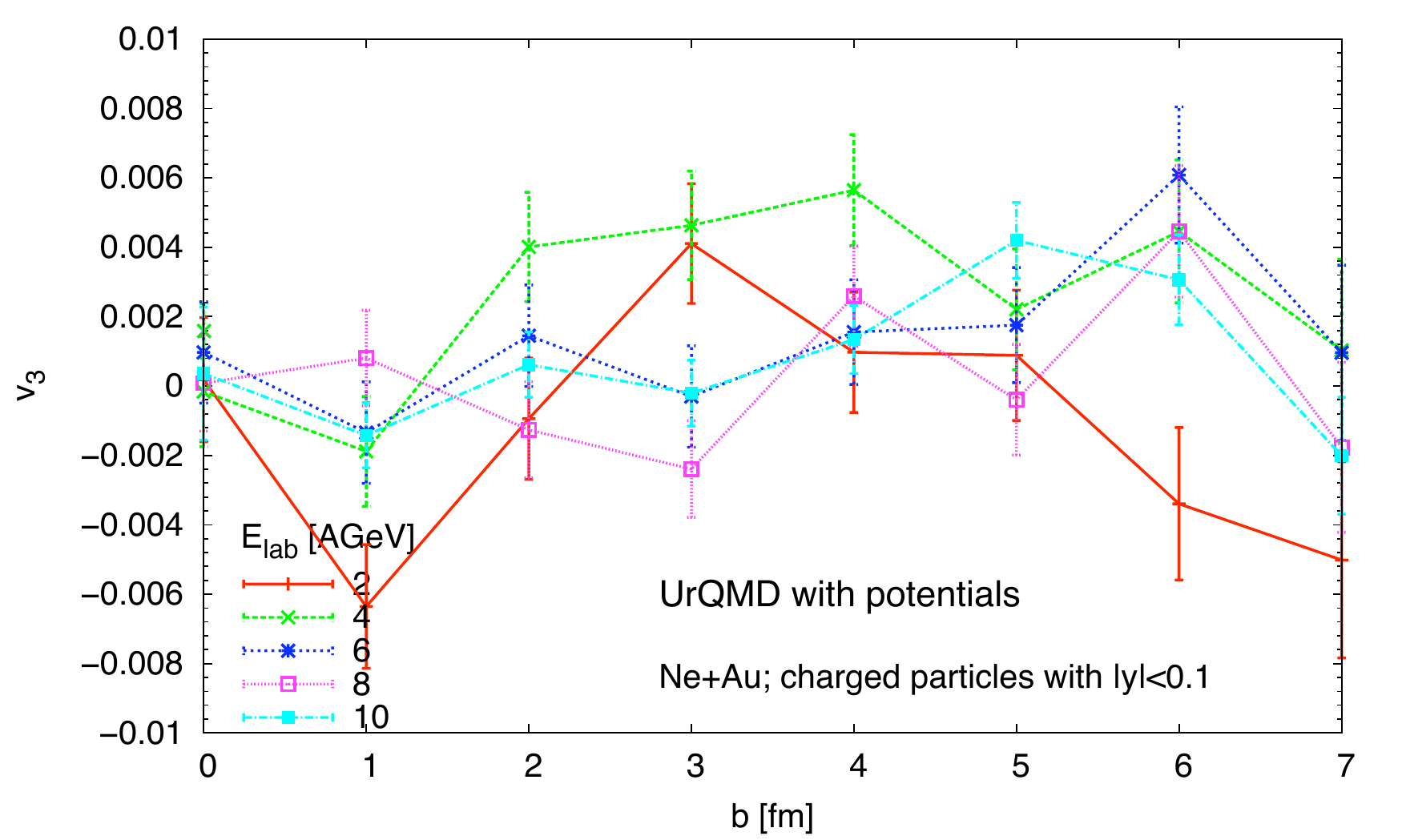,height=5.7cm,width=8.4cm}
    \hspace*{-0.5cm}
    \psfig{figure=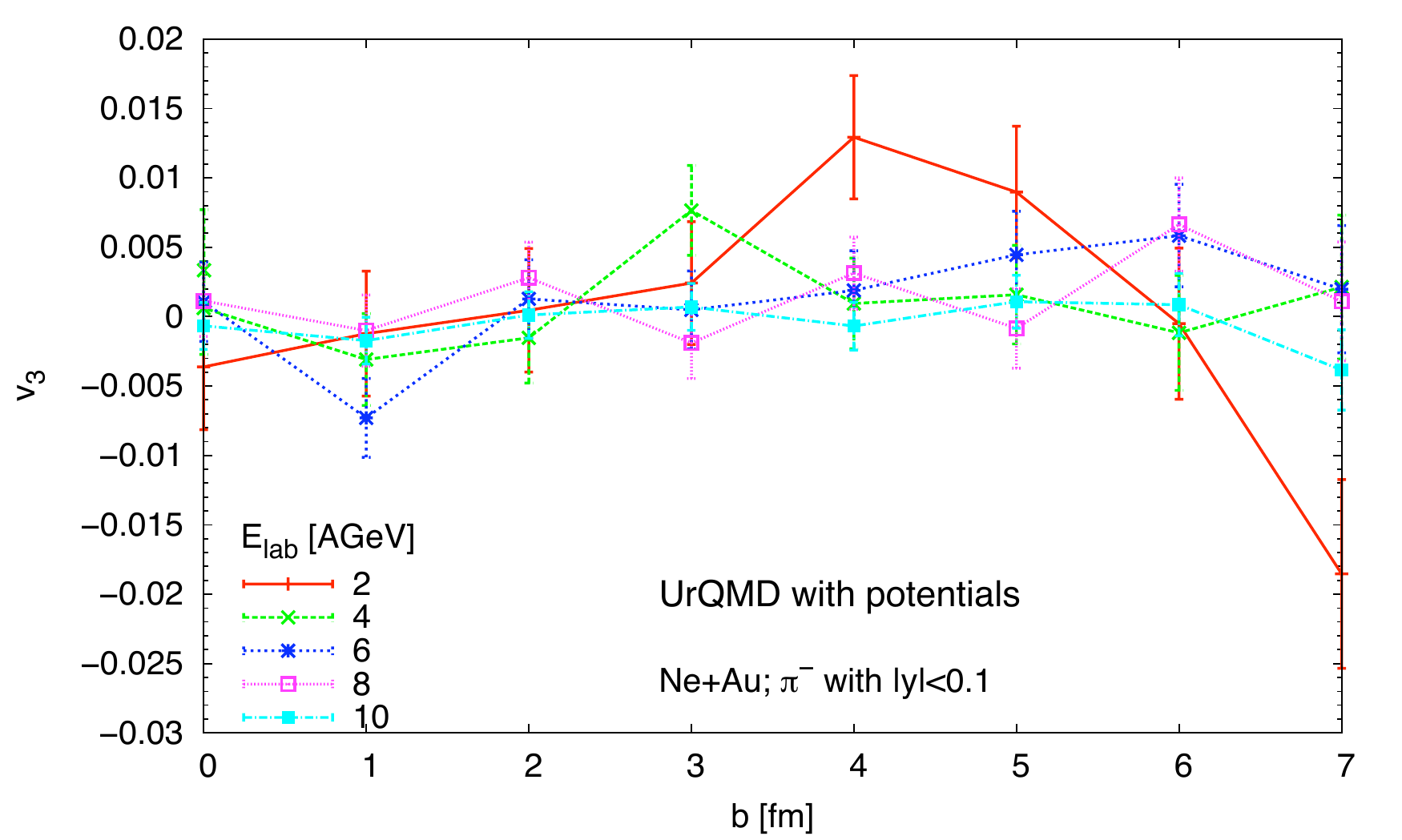,height=5.7cm,width=8.4cm} }

  \vspace*{-.05cm}
  
  {\bf Fig.~5.}  Energy and centrality dependence of the $v_3^{AS}$
  coefficient found by the UrQMD model with potentials for charged
  particles (left panel) and for negative pions (right panel) in the
  ANC Ne+Au.

\end{figure}\label{fig5}

{\bf 3. The dense spot effect on flow patterns.}  The existence of
high density fluctuations in ordinary nuclei is debated for a long
time \cite{Blokhintsev,BohrM} along with their possible applications
in the context of the heavy ion reactions
\cite{Seibert89,Stavinski:WP}.  Here we would like to recapitulate
this issue and to study the influence of the high density fluctuations
on the $v_1^{AS}$, $v_2^{AS}$ and $v_3^{AS}$ coefficients for ANC.  To
estimate this effect, we randomly put 20 nucleons from the Au-nucleus
in a narrow Gaussian distribution inside of the ordinary Monte Carlo
sampled nucleus. This gives us a fluctuating dense spot initial
configuration for the target nucleus. Such a mild assumption is far
from the extremely high densities discussed with respect to flucton
\cite{Blokhintsev,Stavinski:WP}. The dense spot is fixed in the
center of the Au target with an offset in x-direction of half of the
impact parameter between the two colliding nuclei. The condition that
there is an overlap between the dense spot of the radius
1.1\,$B^\frac{1}{3}$~fm and the projectile nucleus of the same radius
is $b - 1.1\,B^\frac{1}{3}\, {\rm fm} \le \frac{b}{2} +
1.1\,B^\frac{1}{3} \, {\rm fm}$. That means, for impact parameters of
the Ne+Au reaction of $b \le 4.4\,B^\frac{1}{3}$~fm $\approx 12$~fm
such an overlap always exists.


\begin{figure}[t]

  \centerline{\psfig{figure=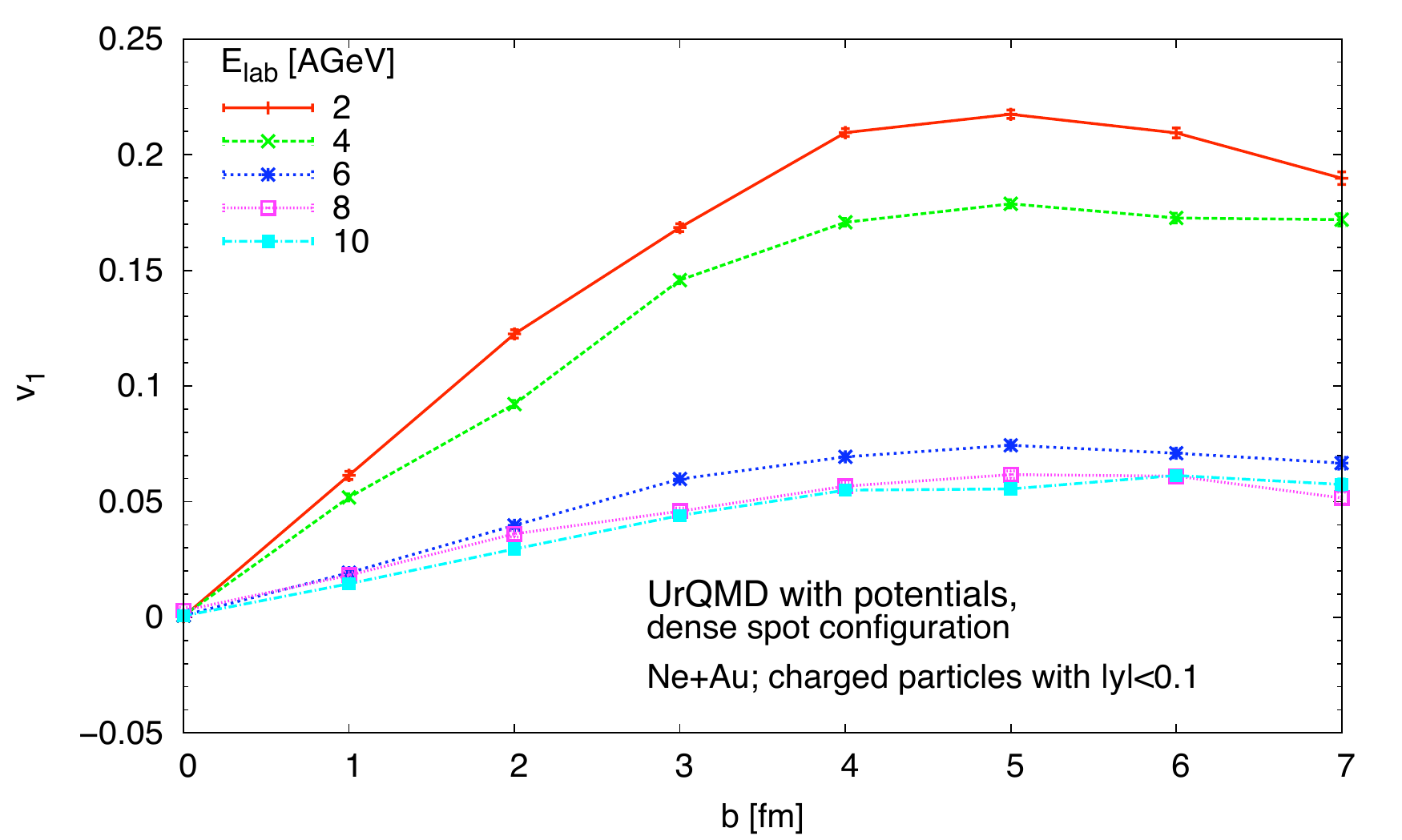,height=5.7cm,width=8.4cm}
    \hspace*{-0.5cm}
    \psfig{figure=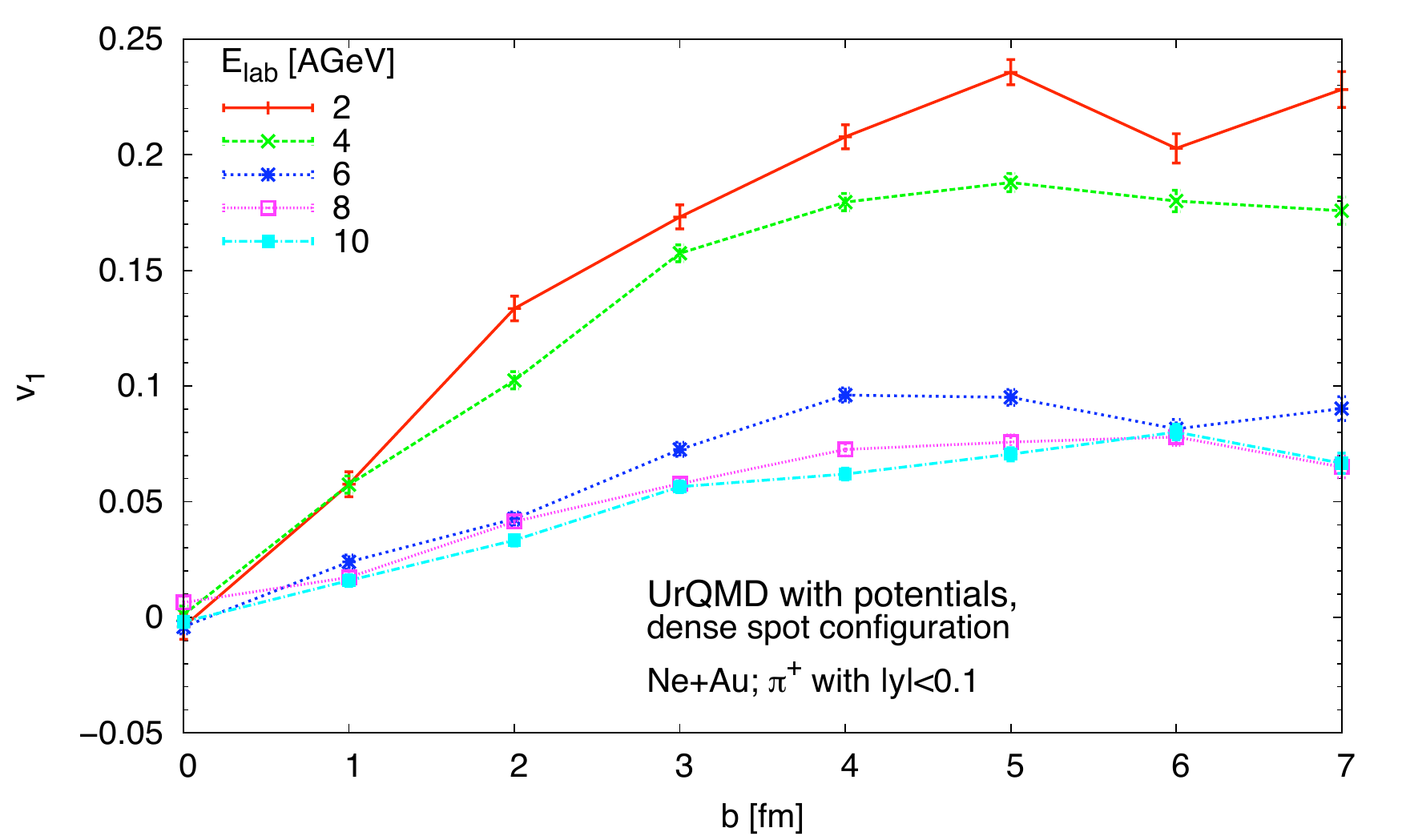,height=5.7cm,width=8.4cm} }

  \vspace*{-.05cm}
  
  {\bf Fig.~6.}  Same as in Fig.~3, but for the dense spot
  configuration (see text for details).
\end{figure}\label{fig6}


\begin{figure}[t]

  \centerline{\psfig{figure=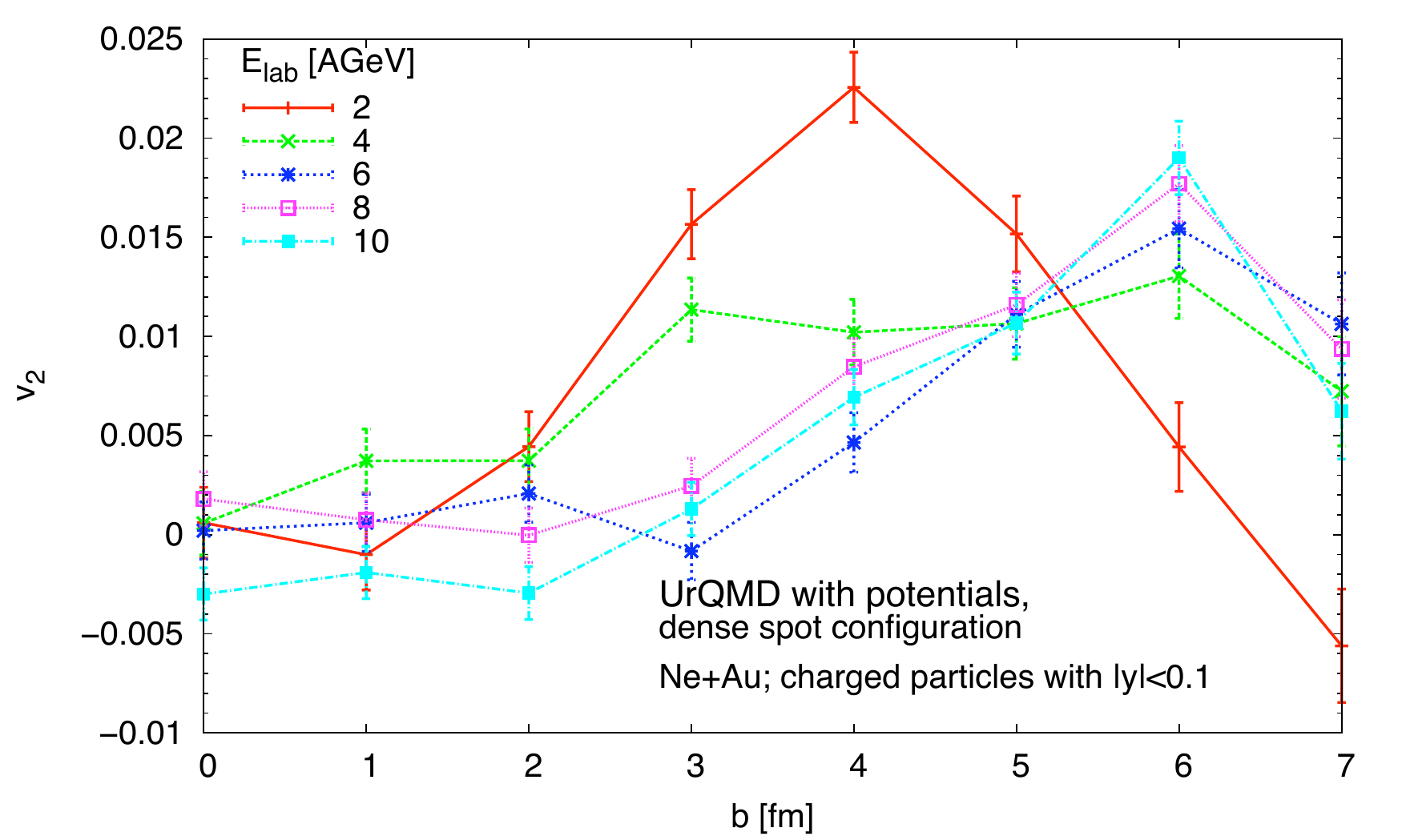,height=5.7cm,width=8.4cm}
    \hspace*{-0.5cm}
    \psfig{figure=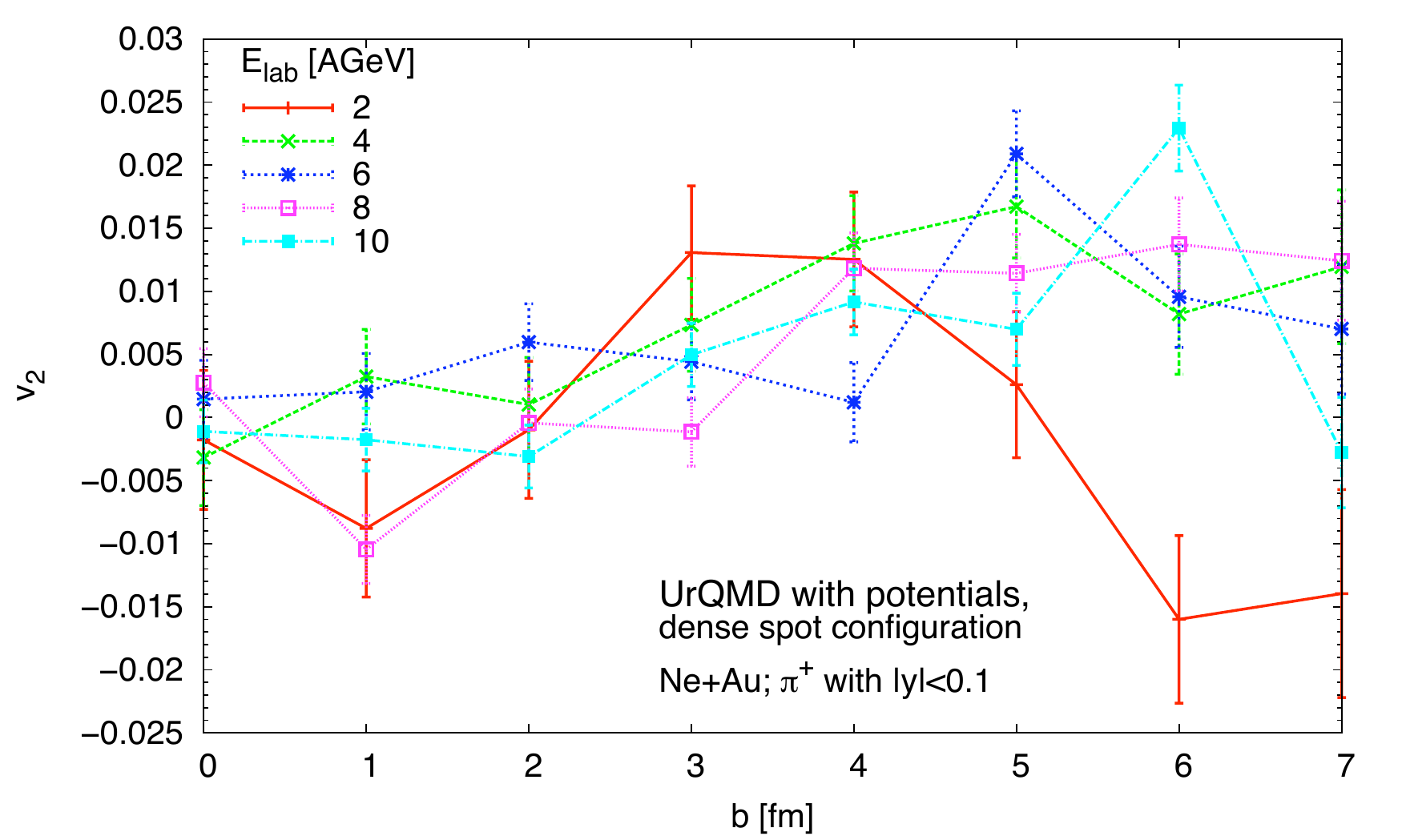,height=5.7cm,width=8.4cm} }

  \vspace*{-.05cm}
  
  {\bf Fig.~7.}  Same as in Fig.~4, but for the dense spot
  configuration.
\end{figure}\label{fig7}

The results for the $v_1^{AS}$, $v_2^{AS}$ and $v_3^{AS}$ coefficients
are presented in Figs.~6--8. Comparing Figs.~3 and 6, one sees that
the directed flow for collisions with the target exhibiting the dense
spot and for collisions without it have almost an identical centrality
dependence for the energies $E_{\rm lab} = 2-4 $ AGeV. However, for
$E_{\rm lab} \ge 5 $~AGeV the $v_1^{AS}$ coefficient with the dense
spot is just 50 \% of the $v_1^{AS}$ coefficient without dense
spot. Thus, there is a non-monotonic behavior of the directed flow
coefficient with respect to the collision energy.  From the analysis
of energy density we conclude that it indicates the change of regimes
from the dominance of the target break up process at $E_{\rm lab} =
2-4$~AGeV to a strong compression and more intense thermalization of
the reaction zone for $E_{\rm lab} \ge 5 $~AGeV.  Qualitatively a
similar picture is valid for the elliptic flow coefficient of the
dense spot configuration: for $E_{\rm lab} = 2-4 $~AGeV the $v_2^{AS}$
coefficients with the dense spot are similar to that ones without it,
whereas for $E_{\rm lab} \ge 5 $~AGeV the elliptic flow coefficient
with the dense spot has different centrality dependence (compare
Figs.~4 and 7).  The latter means that, in contrast to the case of the
dense spot absence, for $E_{\rm lab} \ge 5 $~AGeV the elliptic flow
coefficient of the charged particles is close to zero, i.e.\ $v_2^{AS}
\approx 0$, for $b \le 3$~fm while its peak is now located at $b =
6$~fm.  On the other hand, the $v_2^{AS} $ coefficient of positive
pions exhibits a W-shape for $E_{\rm lab} = 2 $~AGeV, whereas for
$E_{\rm lab} \ge 4$~AGeV its centrality dependence is similar to the
elliptic flow coefficient of protons and charged particles. Also, one
has to admit that for $E_{\rm lab} \ge 4 $~AGeV the maximal amplitudes
of the elliptic flow of both the charged particles, protons and pions
for the dense spot configurations (Fig.~7) are not reduced in strength
compared to the ANC without dense spot (Fig.~4).  Thus, the 50 \%
reduction of the directed flow for the dense spot configurations at
$E_{\rm lab} \ge 4$ does not lead to a dramatic change of the
corresponding elliptic flow at these energies.

Again the essential change in the centrality dependence of the
$v_3^{AS} $ coefficient for the dense spot configuration is seen at
$E_{\rm lab} \le 4 $~AGeV (see Fig.~8): the clear maximum located at
about $b= 3-4$~fm for the no dense spot case (see Fig.~5) disappears in
the dense spot configurations.  Only for energies $E_{\rm lab} \ge 8 $
AGeV one can see some increase of the maximal value of $v_3^{AS} $
coefficient compared to the no dense spot case.


\begin{figure}[t]
  
  \centerline{\psfig{figure=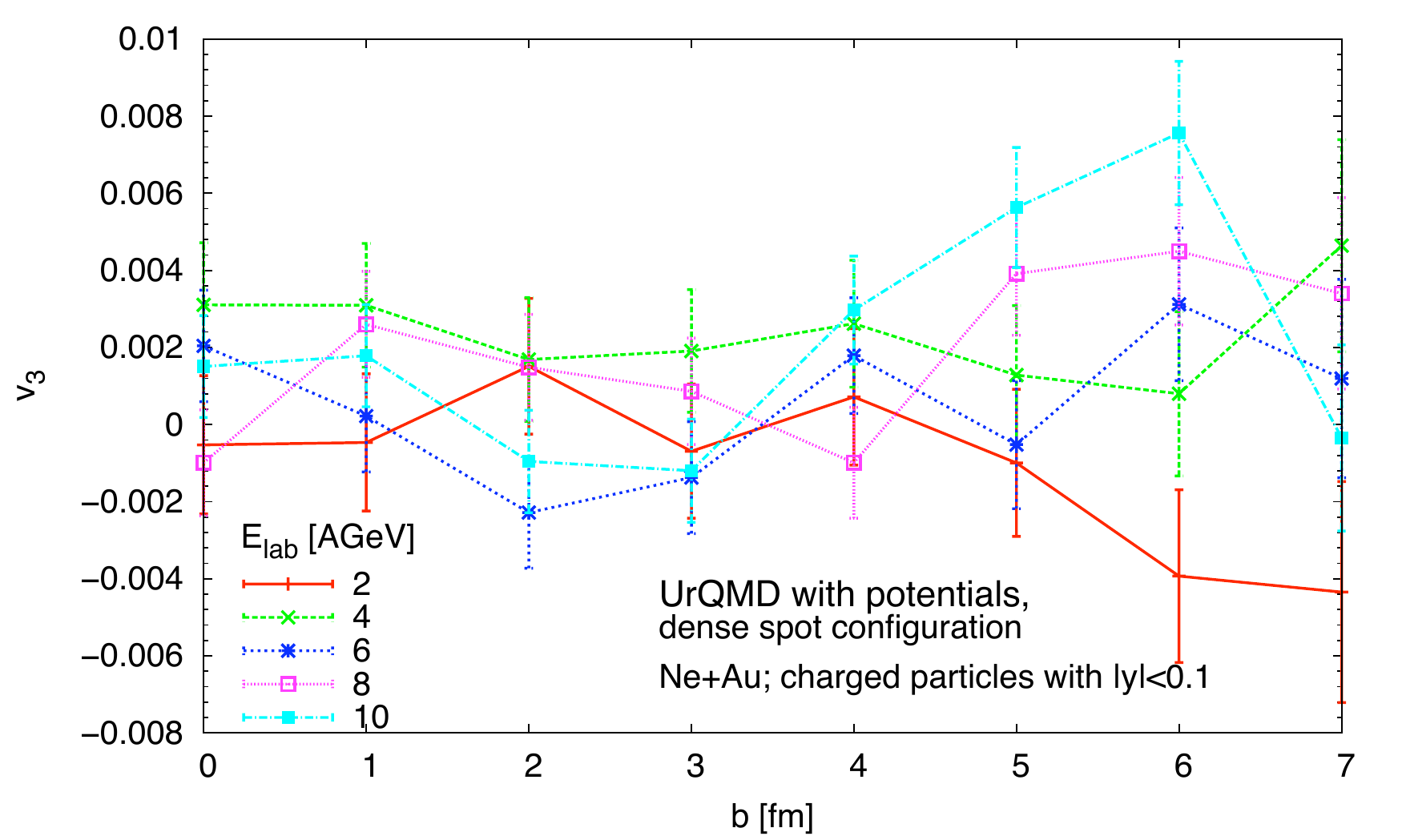,height=5.7cm,width=8.4cm}
    \hspace*{-0.5cm}
    \psfig{figure=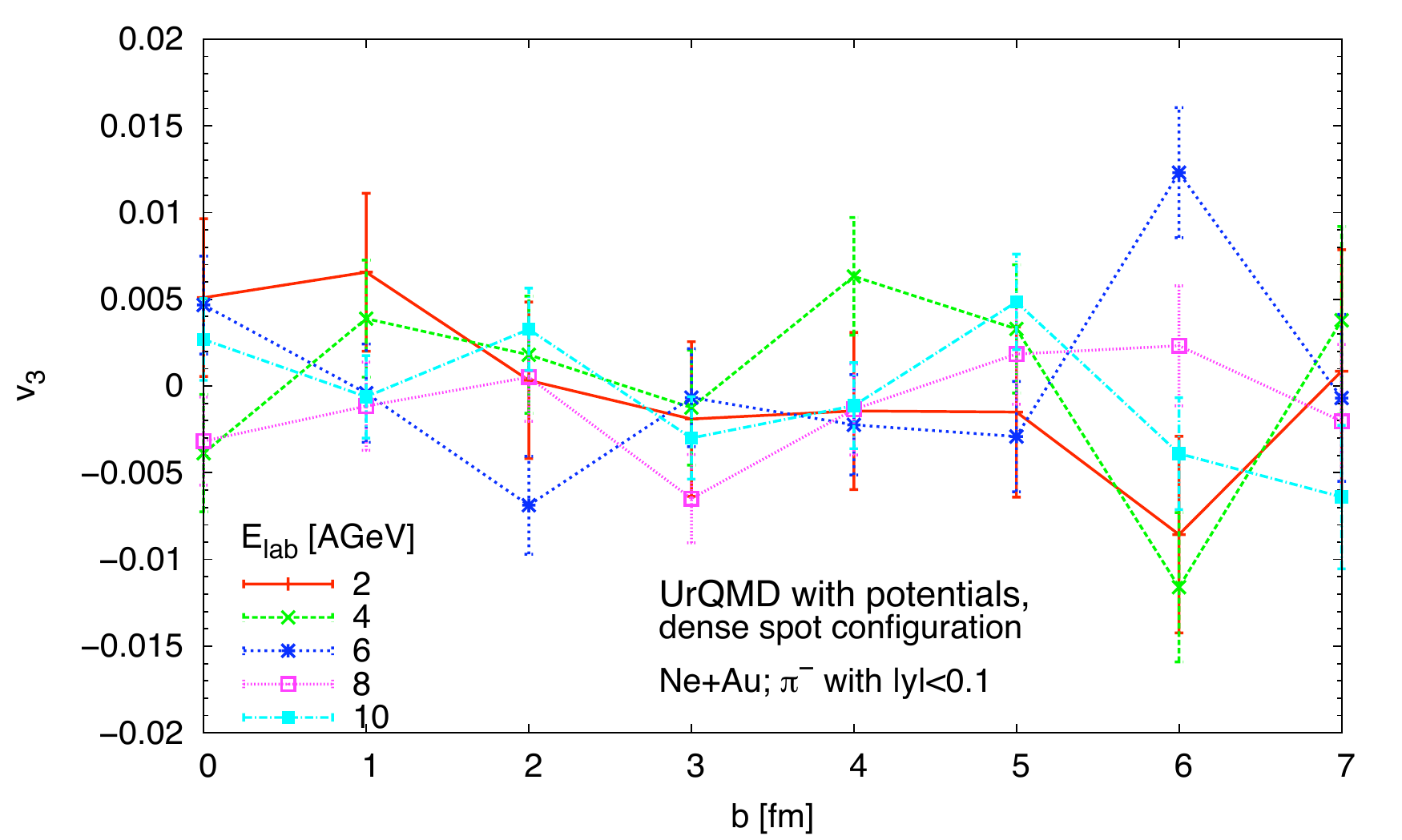,height=5.7cm,width=8.4cm} }

  \vspace*{-.05cm}
  
  {\bf Fig.~8.}  Same as in Fig.~5, but for the dense spot
  configuration.
\end{figure}\label{fig8}

{\bf 4. Conclusions.~} Using the UrQMD model including potentials
\cite{Li:2005gfa,Li:2006ez,Bleicher} here we perform a comprehensive
analysis of the directed, elliptic and triangular flow coefficients
for asymmetric nuclear collisions (ANC) for the first time and suggest
to explore this promising aspect of heavy ion physics in the energy
range of the BMN program at JINR and the FAIR heavy ion program. We
argue that a specific choice of the impact parameter should lead to
the disappearance of the nuclear shadowing effect on the side of
smaller colliding nucleus and show that such an effect leads to an
essential enhancement of the directed and elliptic flows of hadrons in
ANC compared to the symmetric nuclear collisions.

Our analysis shows that the centrality and energy dependencies of the
$v_1^{AS}$, $v_2^{AS}$ and $v_3^{AS}$ coefficients found for ANC are
richer and more complicated compared to symmetric nuclear
collisions. Furthermore, we find that these flow patterns are very
sensitive to the details of the employed interaction which can be used
both for fine tuning of transport codes and for elucidation of the
essential features of the hadron interaction in the medium.

Also, we demonstrate that the possible existence of high density
fluctuations in the large target nucleus may be verified using ANC,
since they allow one to scan the target interior by a smaller
projectile and to draw physical conclusions from the centrality and
the energy dependencies of directed, elliptic and triangular flows on
an event-by-event basis.  The found non-monotonic energy dependence of
the directed flow indicates a higher degree of compression reached in
ANC with the dense spot configuration at lab.\ energies above 5~AGeV.

All these new results allow us to hope that the low energy ANC will
soon become a powerful tool of theoretical and experimental studies of
the strongly interacting matter properties.

\bigskip

{\bf Acknowledgments.}  The authors are thankful to S. Voloshin for
many fruitful discussions.  K.A.B.\ acknowledges the partial support
of the Program 'Fundamental Properties of Physical Systems under
Extreme Conditions' launched by the Section of Physics and Astronomy
of National Academy of Sciences of Ukraine.  The work of A.S.S.\ was
supported in part by the Russian Foundation for Basic Research, Grant
No. 11-02-01538-a. The work of P.R.\ and J.S.\ was supported by BMBF,
HGS-HIRe, and the Hessian LOEWE initiative through the Helmholtz
International Center for FAIR. All computational resources were
provided by the Frankfurt LOEWE Center for Scientific Computing
(LOEWE-CSC).



\begin{thebibliography}{0}

\bibitem{QM11:Phenix}
%
S. Bathe for the PHENIX Collaboration, Plenary talk at the conference ``Quark Matter 2011",
Annecy, France, May 23-28, 2010.
 
\bibitem{QM11:Star}
%
H. Masui for the STAR Collaboration, Plenary talk at the conference ``Quark Matter 2011",
Annecy, France, May 23-28, 2010.


\bibitem{QM11:ALICE}
%
J. Schukraft for the ALICE
Collaboration, Plenary talk at the conference ``Quark Matter 2011",
Annecy, France, May 23-28, 2010.

\bibitem{QM11:ATLAS}
%
P. Steinberg for the ATLAS Collaboration, Plenary talk at the conference ``Quark Matter 2011", Annecy, France, May 23-28, 2010.

\bibitem{QM11:CMS}
%
B. Wyslouch for the CMS Collaboration, Plenary talk at the conference ``Quark Matter 2011",
Annecy, France, May 23-28, 2010.

\bibitem{Voloshin:96}
%
S. Voloshin and Y. Zhang, Z. Phys. C {\bf  70}, 665 (1996).

\bibitem{NICA}
%
A.N. Sissakian, A.S. Sorin, M.K. Suleymanov, V.D. Toneev and G.M. Zinovjev,
Phys. Part. Nucl. Lett. {\bf 5}, (2008) 1.

\bibitem{WhitePaper}
%
NICA White Paper v.3.3, Chapter I,
http://theor.jinr.ru/twiki/pub/
NICA/WebHome/ and refences therein.

\bibitem{Bass:1998ca}
  S.~A.~Bass {\it et al.},
  Prog.\ Part.\ Nucl.\ Phys.\  {\bf 41}, 255 (1998).

\bibitem{Bleicher:1999xi}
  M.~Bleicher {\it et al.},
  J.\ Phys.\ G {\bf 25} (1999) 1859.


\bibitem{Baumgardt:1975qv} 
  %
  H.~G.~Baumgardt et al.,
  Z.\ Phys.\  {\bf A273 } (1975)  359-371.

\bibitem{ANC:1}
%
 J. Gosset et al., Phys. Rev. Lett. {\bf 62}, (1989) 1251. 

\bibitem{ANC:2}
%
B. Adyasevich et al., Nucl. Phys. B (Proc. Suppl.) {\bf 16}, (1990)
419c.
 
\bibitem{ConicWave:10}
%
P.~Rau, J.~Steinheimer, B.~Betz, H.~Petersen, M.~Bleicher and H.~St\"ocker,
  arXiv:1003.1232 [nucl-th].

\bibitem{Randrup:WP}
%
J. Randrup,
NICA White Paper v.3.3,
http://theor.jinr.ru/twiki/pub/
NICA/ WebHome/

\bibitem{Quarkyonic0}
%
for a discussion and estimates see
D. Blaschke, F. Sandin, V. Skokov and S. Typel,
NICA White Paper v.3.3,
http://theor.jinr.ru/twiki/pub/
NICA/WebHome/ and refences therein.




\bibitem{Quarkyonic1}
%
L. McLerran and R. D. Pisarski, Nucl. Phys. A {\bf 796}, (2007) 83.

\bibitem{Quarkyonic2}
%
Y. Hidaka, L. D. McLerran, and R. D. Pisarski, Nucl. Phys. A  {\bf 808}, (2008) 117;
L. McLerran, K. Redlich, and C. Sasaki, Phys. A {\bf 824}, (2009) 86;
K. Fukushima, Phys. Rev. D {\bf 77}, (2008) 114028;
L. Y. Glozman and R. F. Wagenbrunn, Phys. Rev. D {\bf 77}, (2008) 054027.


\bibitem{Quarkyonic3}
%
A. Andronic et al., arXiv:0911.4806v3 [hep-ph] and references therein.

\bibitem{Hofmann:1976dy}
  J.~Hofmann, H.~Stoecker, U.~W.~Heinz, W.~Scheid, W.~Greiner,
  Phys.\ Rev.\ Lett.\  {\bf 36 } (1976)  88-91.


\bibitem{ANC:T79}
%
H. St\"ocker, J. A. Maruhn and W. Greiner,
Z. Phys. A {\bf 293}, (1979) 173.

\bibitem{ANC:T80}
%
H. St\"ocker, J. A. Maruhn and W. Greiner,
Phys.  Rev. Lett. {\bf 44},  (1980) 725.

\bibitem{Stocker:1981zz}
  H.~St\"ocker, C.~Riedel, Y.~Yariv, L.~P.~Csernai, G.~Buchwald, G.~Graebner, J.~A.~Maruhn, W.~Greiner {\it et al.},
  Phys.\ Rev.\ Lett.\  {\bf 47 } (1981)  1807-1810.



\bibitem{Residorf:97}
%
W. Reisdorf and H. G. Ritter,
  Ann.\ Rev.\ Nucl.\ Part.\ Sci.\  {\bf 47}, (1997)  663.
 
 
\bibitem{Ollitrault} 
%
 J.~Y.~Ollitrault,
  Nucl.\ Phys.\  A {\bf 638}, (1998) 195
and references therein. 

\bibitem{Herrmann} 
%
N. Herrmann, J. P. Wessels and Th.  Wienold,
Ann. Rev. Nucl. Part. Sci. {\bf 49}, (1999) 581.

\bibitem{Stock:05}
%
see
R. Stock, J. Phys.G {\bf 30}, (2004) S633 and references therein.

\bibitem{V2scaling}
%
S. S. Adler et al. (PHENIX Collaboration), Phys. Rev. Lett. {\bf 91},  (2003)
182301;
J. Adams et al. (STAR Collaboration),
Phys. Rev. Lett. {\bf  95}, (2005) 122301;
A. Adare  
et al. (PHENIX Collaboration), Phys. Rev. Lett. {\bf 98}, (2007) 162301. 


\bibitem{Alver:10}
%
B. Alver and G. Roland,
Phys. Rev.  C {\bf 81}, (2010) 054905.


\bibitem{AntiFlow:Bravina}
%
for a discussion see
E. E. Zabrodin, C. Fuchs, L. V. Bravina and A.  Faessler,
Phys. Rev. C  {\bf 63}, (2001) 034902.

\bibitem{FlowP}
%
K. G. R. Doss et al., Phys. Rev. Lett. {\bf 57}, (1986) 302. 

\bibitem{SNC:3}
%
Ph. Crochet et al., FOPI Collaboration, Nucl. Phys. A {\bf 627}, (1997) 522.

\bibitem{Stoecker:2004qu}%
  H.~St\"ocker,
  Nucl.\ Phys.\  {\bf A750 } (2005)  121-147.


\bibitem{AntiFlow:SIS}
%
A. Wagner et al., Phys. Rev. Lett.  {\bf 85}, 18 (2000).

\bibitem{AntiFlow:Bertsch}
%
B.-A. Li, W. Bauer, and G. F. Bertsch, Phys. Rev. C {\bf 44}, 450 (1991).


\bibitem{AntiFlow:Horst}
%
S. A. Bass, C. Hartnack, H. St\"ocker, and W. Greiner, Phys. Rev. Lett. {\bf 71}, 1144 (1993); Phys. Rev. C {\bf 50}, 2167 (1994).

\bibitem{Li:2005gfa}
  Q.~-f.~Li, Z.~-x.~Li, S.~Soff, M.~Bleicher, H.~Stoecker,
  J.\ Phys.\ G {\bf G32 } (2006)  151-164.


\bibitem{Li:2006ez}
  Q.~-f.~Li, Z.~-x.~Li, S.~Soff, M.~Bleicher, H.~Stoecker,
  J.\ Phys.\ G {\bf G32 } (2006)  407-416.

\bibitem{Bleicher}
%
H. Petersen, Q. Li, X.  Zhu  and M.  Bleicher, 
Phys. Rev. C {\bf 74}, 064908 (2006).



\bibitem{Ollitrault:v2}
%
J.-Y. Ollitrault, Phys. Rev. D {\bf 46},  (1992) 229.

\bibitem{Bass:10}
%
H. Petersen, G.-Y.Qin, S. A. Bass and B. M{\" u}ller,
arXiv:1008.0625v2 [nucl-th].




\bibitem{Blokhintsev}
%
D. I. Blokhintsev, ZhETP {\bf 6}, (1958) 995. 

\bibitem{BohrM}
%
A. Bohr and B. Mottelson, "Nuclear Structure", Benjamin Press, New York, 
1975, Vol.2. 


\bibitem{Seibert89}
%
D. Seibert,  Phys. Rev. Lett. {\bf 63},   (1989) 136.

\bibitem{Stavinski:WP}
%
A. Stavinskiy et al., NICA White Paper v.3.3,
http://theor.jinr.ru/twiki/pub/
NICA/WebHome/ and refences therein.




\end{thebibliography}
\end{document}